\documentstyle[aps,epsf,floats,twocolumn,prl] {revtex}
        
\begin{document}
\twocolumn[\hsize\textwidth\columnwidth\hsize\csname @twocolumnfalse\endcsname

\title  {       Exactly solvable model with two conductor-insulator 
  transitions driven by impurities
}
\author {       M. Bauer
\and            O. Golinelli
}
\address{       Cea Saclay, Service de Physique Th\'eorique, 
                91191 Gif-sur-Yvette, France
\\              email: bauer and golinelli@spht.saclay.cea.fr
}
\date {July 3, 2000, revised October 25, 2000 and  January 11, 2001} 

\preprint{      Preprint Spht 00/087 ; cond-mat/0006472
}
\maketitle

\begin{abstract}
We present an exact analysis of two conductor-insulator
transitions in the random graph model. The average connectivity is
related to the concentration of impurities. The adjacency matrix of a
large random graph is used as a hopping Hamiltonian. Its spectrum has
a delta peak at zero  energy. Our analysis is based
on an explicit expression for the height of this peak,
and a detailed description of the localized eigenvectors and of their
contribution to the peak. Starting from the low connectivity (high
impurity density) regime, one encounters an insulator-conductor
transition for average connectivity $1.421529\cdots$ and a
conductor-insulator transition for average connectivity
$3.154985\cdots$. We explain the spectral singularity at average
connectivity $e=2.718281\cdots$ and relate it to another enumerative
problem in random graph theory, the minimal vertex cover problem.
\end{abstract}

\draft



]
\narrowtext

Random graphs have motivated a lot of work both in mathematics and in
physics.  In the random graph model, $N$ points numbered $1,2,\cdots , N$
are used as vertices. A pair of (distinct) vertices $\{i,j\}$ is
connected by an edge with probability $p$ and the edges are
independent. The adjacency matrix of the graph is the symmetric matrix
$H$ with matrix element $H_{i,j}=1$ if vertices $i$ and $j$ are
connected by an edge and zero otherwise. The average connectivity
(i.e. average number of neighbors of a given vertex) is $\alpha=pN$.

An interesting asymptotic regime emerges when $\alpha$ is kept fixed as $N$
goes to infinity. The connectivity serves as a parameter, and several phase
transitions can be observed.

According to the seminal papers on the subject \cite{erdos}, for $\alpha<1$
all connected components are finite, and only trees contribute to the
extensive (i.e. proportional to $N$) quantities but for $\alpha >1$, a
finite fraction of the points lies in a single connected component, the
giant component. So there is a second order (classical) percolation
transition at $\alpha=1$.

The adjacency matrix $H$ can be used as a Hamiltonian that describes
hopping of electrons from one site to another if the two are connected by an
edge. A large average connectivity means small concentration of
impurities and vice-versa.   

The spectrum of $H$ is relevant to many problems in
physics and has been investigated by many authors. It contains
\cite{bauer00a} an infinity of delta peaks for any $\alpha$ and also a
continuous component for large enough $\alpha$. It has been argued
\cite{evangelou92} that for $\alpha \simeq 1.4$, a quantum percolation
transition occurs. This means that the structure of eigenvectors changes :
below this value, all eigenvectors are localized, but above, some
eigenvectors occupy a finite fraction of the system. It is believed
\cite{evangelou92} that the continuous component in the spectrum appears at
the same threshold. Also an anomaly in the spectrum near the energy $0$ for
$\alpha \simeq 2.7$ has been noticed in \cite{bauer00a}.

The main results presented in this letter are an analytical
expression for $z(\alpha)$, the average height of the delta peak
at  zero energy in the spectrum of $H$ in the thermodynamic
limit, and the identification of the corresponding eigenvectors, 
leading to precise predictions for two quantum percolation transitions in
this model : a delocalization (insulator-conductor) transition at
$\alpha_d=1.421529\cdots$ and a relocalization (conductor-insulator)
transition\cite{remark1} at $\alpha_r=3.154985\cdots$. Surprisingly,
$z(\alpha)$ is analytic at the classical and quantum 
percolation transitions and non-analytic only at $\alpha=e$. 
 
The random graph model displays some mean field $D=\infty$ features because the
neighbors of a site are chosen with equal probability among all
the other points, with the consequence that loops are large. This is
in strong contrast with real materials where 
some underlying geometry (e.g. a lattice) restricts
the possible neighbors of a point, even before
impurities destroy bonds. But once the random graph is chosen,
each point has $\alpha$ neighbors on average so it interacts
effectively with only a finite number of points, and this is far from
mean field. In the conducting phase, the delocalized
states live on the giant component, whose effective average
connectivity varies between $2.092917\cdots$ and $3.312453\cdots$.
Values of this order can be achieved 
in real two or three dimensional disordered systems, and this
suggests that analogous insulator-conductor-insulator transitions
might be observed in real materials. Indeed, the random graph model
 gives a rich phase diagram with a single
parameter, and some real systems should exhibit at least the same
complexity. On the other hand, we do not expect that the precise
values of $\alpha_r$ and $\alpha_d$ will appear in real
materials. Whether the critical exponents we obtain
(which look very much like mean field exponents) remain valid above some upper
critical dimension, and whether this covers some physical situation
is still unclear.     
   
\bigskip

The explicit formul\ae\ below will involve five functions which are
analytic for real positive $\alpha$ except maybe at three special
values. These functions are generating functions
related to the enumeration of various types of trees. In particular,
they all have the same small $\alpha$ expansion.
We start with a short description of these functions. 

The most basic generating function for tree enumeration is the Lambert
function $W(\alpha)$, which is analytic on the real positive axis and
satisfies a fixed point equation,
\begin{equation}
  \label{w}
 W=\alpha e^{-W}.
\end{equation}
This implies that $W=-\sum_{n\geq 1} (-\alpha)^n\frac{n^{n-1}}{n!}$,
and $n^{n-1}$ is the number of labeled rooted trees on $n$ vertices.

However, as a fixed point, $W$ is unstable for $\alpha >e=\exp (1)$,
and there is a stable periodic orbit of length 2. This leads us to introduce 
the real functions $A(\alpha)$ and $B(\alpha)$
solving the symmetric system
\begin{equation}
  \label{w+-}
  A=\alpha e^{-B}, \qquad  B=\alpha e^{-A}.
\end{equation}
and satisfying the condition that $A < B$ for $\alpha >e$. For
$0\leq \alpha \leq e$, there is only one solution to the system (so no
condition is needed in this range of $\alpha$'s), namely
$A(\alpha)=B(\alpha)=W(\alpha)$. Thus $e$ is a special value where a
branch point occurs. We shall see later that 
$A$ and $B$ can also be viewed as generating functions for certain bicolored
trees, with $-\alpha$ as weight for vertices of one color and $-\alpha
e^{-\alpha}$ for vertices of the other color. This interpretation is
relevant at large $\alpha$.

We shall need one more function, $A^{\star}(\alpha)$ which is the smallest
solution of the equation
\begin{equation}
  \label{w-bar}   
  A^{\star}=A e^{\alpha (e^{A^{\star}-A}-1)}.
\end{equation}
Clearly $A^{\star}=A$ is a solution, but
in general not the smallest. The $\alpha$'s for which the
solutions of (\ref{w-bar}) exhibit a branch point are such that
$A\alpha=1$ which leads to   
\begin{equation}
  \label{al+-}
  2\log \alpha = \alpha e^{-1/\alpha}.
\end{equation}
This equation has two solutions, $\alpha_d=1.421529\cdots$, and
$\alpha_r=3.154985\cdots$. In fact, $\alpha_d$ is the solution of
the simpler equation $ 2\alpha \log \alpha = 1$. In the interval
$]\alpha_d,\alpha_r[$, $A^{\star} < A$, but outside this interval
$A^{\star}=A$. The function $A^{\star}$ , which coincides with $A$
for large and small $\alpha$, and thus shares the same combinatorial
interpretations in these regimes, appears in this discussion via the
explicit evaluation of the sum (\ref{sum}).   

Replacing $A$'s by $B$'s in (\ref{w-bar}) leads to another function
$B^{\star}(\alpha)$. The functions $W$, $A$, $B$, $A^{\star}$ and $B^{\star}$
are plotted in Fig.~1.
\begin{figure}
  \centering\leavevmode
  \epsfxsize=8.5truecm \epsfbox{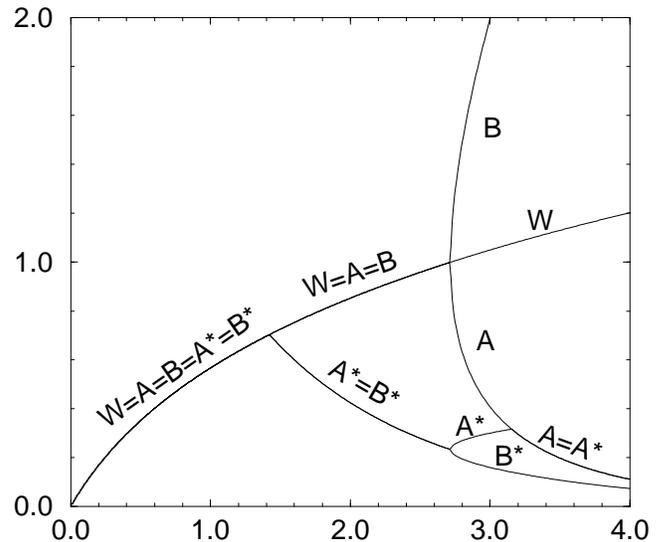}
  \caption{\em The special functions $W$, $A$, $B$, $A^{\star}$ and 
  $B^{\star}$ with their domains of coincidence.The branchings occur
  at the special values $\alpha_d,e$ and $\alpha_r$.}
  \label{f1}
\end{figure}

\bigskip

We now argue that $z(\alpha)$, the height of the delta peak at the
eigenvalue zero in the spectrum of an infinite random adjacency matrix,
is given by the explicit formula 
\begin{equation}
  \label{zero}
  z(\alpha)=-1+\frac{A +B + A B}{\alpha}.
\end{equation}

What we have proved rigorously \cite{bauer00b} is that (\ref{zero}) is
true for $\alpha \leq e$ and that the r.h.s of (\ref{zero}) is an
upper bound for $z(\alpha)$ for $\alpha >e$. Here is the idea. 
Let $Z(G)$ denote the number of zero eigenvalues of the
adjacency matrix of the graph $G$. Suppose $G$ has a vertex
$v$ with exactly one neighbor, say $v'$ (so $v$ is a leaf of $G$).
Leaf removal consist in removing from $G$ the vertices $v$ and $v'$
and all edges that meet those two vertices, leading to a new graph
$G'$. In \cite{bauer00} it is shown that $Z(G)=Z(G')$. Iteration of
leaf removal leads to a graph with consists of say $I$ isolated points
and a subgraph $C$ 
(called the core of $G$) without leaves or isolated points.
Then $Z(G)=I+Z(C)$. In \cite{bauer00b} it is proved that if $G$ is
a random graph of size $N$ and average connectivity $\alpha$, the size
of $C$ is $o(N)$ if \cite{remark2} and only if $\alpha \leq e$ and that
$I/N=-1+\frac{A +B + A B}{\alpha}+o(1)$ for large $N$. But almost by
definition, $Z(G)/N=z(\alpha)+o(1)$, and this completes the proof. 

Then  $z(\alpha)$ is singular at $\alpha=e$. It coincides with
the analytic function
\begin{equation}
z_{an}(\alpha) \equiv -1+\frac{2W+W^2}{\alpha}
\end{equation}
for $\alpha \leq e$  but $z(\alpha)>z_{an}(\alpha)$ for
$\alpha >e$. 

We believe that (\ref{zero}) is true as an equality even for $\alpha >e$ 
for two reasons. First, the Monte Carlo simulations in Fig.~2 are
in perfect agreement with (\ref{zero}) for all $\alpha$'s. Second,   
from combinatorial arguments \cite{bauer00a,bauer00}, one can show
that $z(\alpha)$ can be expressed formally as
\begin{equation} 
  \label{decomp}
  z(\alpha) = \sum_T \frac{1}{Aut(T)}(-\alpha)^{E(T)},
\end{equation}
where the sum over $T$ is over isomorphism classes of bicolored (say brown
and green) trees with at least one green vertex, $Aut(T)$ is the size of
the automorphism group of $T$ and $E(T)$ the number of edges of $T$.  In
fact, (\ref{decomp}) is an identity of analytic functions for $\alpha \leq
1/e$.
For larger $\alpha$'s, the sum in (\ref{decomp}) is divergent, but a
natural partial resummation turns it into a convergent expression as follows.
Any bicolored tree can be obtained from a bicolored tree
with only green leaves by appending brown leaves at the green
vertices, so we
can explicitly sum over all brown leaves in the sum over bicolored trees.
This yields an identity valid at small $\alpha$:
\begin{equation}
  \label{bicolsum}
  z(\alpha)=\sum_T \frac{1}{Aut(T)} (-\alpha)^{V_B(T)}
  \alpha ^{V_G(T)-1} e^{-V_G(T)\alpha},
\end{equation}
where the sum over $T$ is over bicolored trees with only green leaves,
and where $V_B(T)$ and $V_G(T)$ are, respectively, the number of brown 
and green
vertices of $T$. Regularizing the sum by restricting $T$ to have 
diameter at most $d$ leads to a recursive sequence depending on
$d$ whith limit the r.h.s of (\ref{zero}).
Moreover, the sum in (\ref{bicolsum}) can be regrouped as
\begin{equation}
  \label{groupbicolsum}
  z(\alpha)=\sum_{n \geq 1} \frac{\alpha^{n-1} e^{-n\alpha}}{n!}
  \frac{S_{n-1}(-n\alpha)}{n},
\end{equation}
where the Stirling polynomials $S_n$ are defined by:
 \begin{equation}
  \label{stir}
  e^{x(e^t-1)}=\sum_{n\geq 0} S_n(x) \frac{t^n}{n!}.
\end{equation}
The sum in (\ref{groupbicolsum}) converges to the r.h.s. of
(\ref{zero}) for all $\alpha$'s, giving further evidence for the
validity of (\ref{zero}).  

The appearance of bicolored trees with only green leaves is no
hazard. Consider the following pattern : a subgraph of the random graph
which is a finite tree $T$ and which is such that one of its two
bicolorings has only green leaves and the green vertices share no edges
with the complement of $T$ in the random graph. One can show that the
frequency of apparition of a maximal (that is, not contained in a larger
tree with the same properties) such tree in the random graph is
\cite{remark1bis}
\begin{equation}
  \label{weight}
  \frac{1}{Aut(T)} B^{V_B(T)}\alpha^{V_G(T)-1} e^{-V_G(T)\alpha},
\end{equation}
and that exactly $V_G(T)-V_B(T)$ eigenvectors with eigenvalue $0$ of the
adjacency matrix of the random graph are localized on $T$ (in fact, they
are even localized on the green vertices of $T$). The sum
\begin{equation}
  \label{sum} 
  z_{loc}(\alpha)\equiv \sum_T \frac{V_G(T)-V_B(T)}{Aut(T)} 
  B^{V_B(T)}\alpha^{V_G(T)-1} e^{-V_G(T)\alpha}
\end{equation}
of these non-negative contributions gives a lower bound for
$z(\alpha)$. Note the striking similarity between the sums defining
$z(\alpha)$ and $z_{loc}(\alpha)$. Explicitly, 
\begin{equation}
  \label{zfin}
  z_{loc}(\alpha) = -\frac{Be^{A^{\star}}}{\alpha}+
  \frac{A^{\star} +B + A^{\star}  B}{\alpha},
\end{equation}
and one checks that
\begin{equation}
  \label{inegal}
  \begin{array}{ll} 
    z_{loc}(\alpha)=z(\alpha) & \mbox{for $0 \leq
    \alpha \leq \alpha_d$ and $\alpha \geq \alpha_r$}, \\
    z_{loc}(\alpha) < z(\alpha) & \mbox{for $\alpha_d < \alpha < \alpha_r$}.  
  \end{array}
\end{equation}

This is our second important result : for $0 \leq \alpha \leq \alpha_d$ and
$\alpha \geq \alpha_r$, all the (extensive) contributions to the kernel of
random adjacency matrices come from vectors localized on finite bicolored
trees, with green leaves only, attached to the rest of the random graph
only by brown vertices. However, this is not true for $\alpha_d
< \alpha < \alpha_r$, where $z_{loc}(\alpha)<z(\alpha)$.

Moreover, at the ``critical'' values $\alpha_d$ and $\alpha_r$, the
distribution of $V_B(T)$, $V_G(T)$, or even
$V_G(T)-V_B(T)$ becomes large : the second moment diverges. For example,
\begin{equation}
  \label{average1}
  \langle V_G(T)-V_B(T) \rangle =\frac{1+2\alpha_d-\alpha_d^3}
  {\alpha_d^3-\alpha_d^2}=1.139353\cdots  
\end{equation}
at connectivity $\alpha=\alpha_d$ but
\begin{equation}
  \label{average2}
  \langle (V_G(T)-V_B(T))^2 \rangle \sim \frac{\alpha_d^5-\alpha_d^4-3
  \alpha_d^3+\alpha_d^2 +3\alpha_d+1}{(2\alpha_d^4-\alpha_d^3-
  \alpha_d^2)(\alpha_d-\alpha)} 
\end{equation}
when $\alpha$ approaches $\alpha_d$ from below.  The presence of
unbounded fluctuations indicates that infinitely extended objects are
responsible for the difference between $z_{loc}(\alpha)$ and $z(\alpha)$ in
the range $\alpha_d < \alpha < \alpha_r$. Thus we infer that at zero
energy, the eigenvectors of the hopping Hamiltonian exhibit a
delocalization (insulator-conductor) transition at $\alpha_d$ and a
relocalization (conductor-insulator) transition at
$\alpha_r$. The prediction of delocalization at $\alpha_d$ is in numerical
agreement with Monte-Carlo simulations\cite{evangelou92}. It
turns out that the same delocalization value was already found by Harris
\cite{evangelou92,harris} in a loopless model of random Bethe trees. This
is surprising because at $\alpha_d$, the random graph has already
(many) loops.

Note the analog of (\ref{groupbicolsum}) for $z_{loc}(\alpha)$: 
\begin{equation}
  \label{expzfin}
  z_{loc}(\alpha)=\sum_{n\geq 1}\frac{\alpha^{n-1}e^{-n\alpha}}{n!} \;
  \frac{n(1+B)S_{n-1}(nB)-S_n(nB)}{n}.  
\end{equation}

\begin{figure}
  \centering \leavevmode
  \epsfxsize=8.5truecm \epsfbox{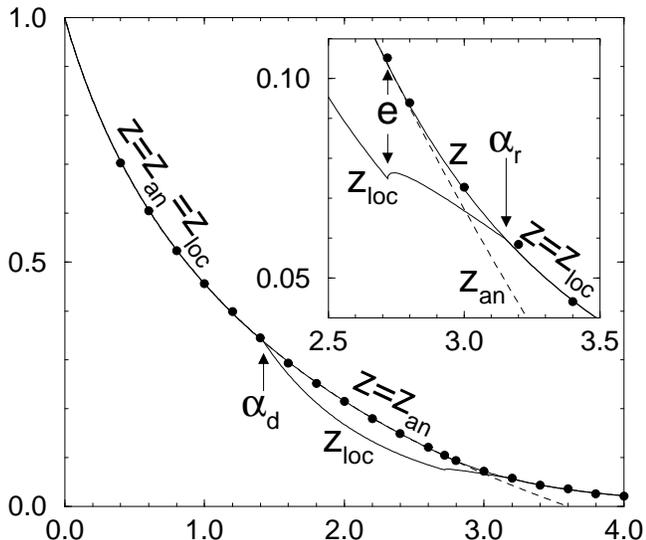}
  \caption{\em Summary
  of the results. The functions $z$ and $z_{loc}$ correspond to full lines,
  whereas $z_{an}$ corresponds to the dashed line. The inset
  magnifies the region $2.5 \leq \alpha \leq 3.5$. The height of the delta
  peak $z$ is to be compared with Monte-Carlo simulations, represented as
  black dots. The error bars are of order $2.10^{-3}$, much smaller than
  the symbols on the main plot, and of the order of the size of the symbols
  in the inset. The negative values of $z_{an}$ are not represented.}
  \label{f2}
\end{figure}

Our main results are summarized in Fig.~2. 

\bigskip

As noticed before, the expressions for $z(\alpha)$, $z_{loc}(\alpha)$ and
their difference show a singularity at $\alpha=e$, due to the fact 
that the frequency of certain patterns is non-analytic. It is tempting
to relate this singularity to the flattening of the spectral
distribution near the zero eigenvalue for $\alpha \simeq 2.7$ observed
in \cite{bauer00a}.  In
fact, a transition in properties of random graphs at $\alpha=e$ has already
been observed in a different context, minimal vertex covers
\cite{weigt,hartmann}. A vertex cover of a graph is a subset of the
vertices containing at least one extremity of every edge of the graph. An
analogous concept is that of an edge disjoint system, i.e. a subset of the
edges such that no two edges in the subset have a vertex in common. Given a
graph $G$ on $N$ vertices, let $X(G)$ be the minimal size of a vertex
cover and $Y(G)$ be the maximal size of an edge disjoint system. It is
clear that $X(G)\leq N$ and $2Y(G)\leq N$. Those quantities share a very
important property with $Z(G)$, they behave simply under leaf removal 
\cite{bauer00b} : $X(G')=X(G)-1$, $Y(G')=Y(G)-1$ whereas
$Z(G')=Z(G)$. For general graphs, $X(G)\geq
Y(G)$ and $Z(G)\geq N-2X(G)$. Examples show that
$Z(G)-N+2Y(G)$ can have any sign.

But if  $G$ is a random graph with $\alpha \leq e$, leaf removal leaves a
core of size $o(N)$ so if
$x(\alpha)$, $y(\alpha)$ and $z(\alpha)$ denote the limits of the averages
of $X(G)/N$, $Y(G)/N$ and $Z(G)/N$ when $N \rightarrow \infty$ we get
\begin{equation}
\label{guard}
x(\alpha)=y(\alpha)=\frac{1}{2}(1-z(\alpha)) \quad \mathrm{for} \quad
\alpha \leq e. 
\end{equation}

This is the formula for $x(\alpha)$ when $\alpha \leq e$
obtained by Hartmann and Weigt via a replica symmetric ansatz
\cite{weigt,hartmann}. However, the relation 
(\ref{guard}) has to break down at some $\alpha > e$,
because for large $\alpha$, $z(\alpha)$ is exponentially small in
$\alpha$, whereas a rigorous result \cite{frieze} predicts that
\begin{equation}
  \label{fr}
  x(\alpha)=1-\frac{2}{\alpha}(\log \alpha- \log \log \alpha -\log 2
  +1)+o(\frac{1}{\alpha}). 
\end{equation}
Simulations seem to indicate that the simple relation between $x(\alpha)$ and
$z(\alpha)$ breaks down at $e$, together with replica
symmetry stability. Our interpretation is that the remnant after
leaf removal is a complicated graph with a size of order $N$ which can
be described only by a more refined replica asymmetric order parameter. 

\bigskip

In this paper, we have given an exact description of the size of the kernel
of the hopping Hamiltonian associated to a large random graph and of
the structure of the 
corresponding eigenvectors, leading to predictions of a delocalization and
relocalization transition at $\alpha_d$ and $\alpha_r$. We have also
explained the singularities affecting all spectral and several combinatorial
properties at $\alpha=e$. A surprising fact is that all such
properties are regular at the (classical) percolation transition at
$\alpha=1$. Possibilities to extend these results to more realistic
models remain to be explored. 

\vglue -0.5cm


\begin{references}
\bibitem{remark1} It is generally believed that the first delocalized
  states appear at zero-energy. However, the disappearance of
  delocalized states at zero-energy for $\alpha_r=3.154985\cdots$ does
  not exclude that some exited delocalized states remain above this threshold.
\bibitem{remark2} If $\alpha <1$, it is obvious that $C$ is $o(N)$,
  because $G$ is essentially a collection of trees.
\bibitem{remark1bis} A striking feature of Eq.~(\ref{weight}) is that
  despite the maximality condition, it depends only on the most basic
  features of $T$: the number of green and brown vertices. Each brown
  vertex gives a weight $B$, which is non-analytic at $e$. So, at $e$,
  the random graph has a geometric phase transition in that the
  frequency of certain patterns is non-analytic. This is related to
  the appearance of a large core at the same threshold.
\end{references}
\end{document}